\documentclass[3p,times,twocolumn]{elsarticle}

\usepackage{ecrc}

\volume{00}

\firstpage{1}

\journalname{Nuclear and Particle Physics Proceedings }

\runauth{}

\jid{nppp}

\jnltitlelogo{Nuclear and Particle Physics Proceedings }

\usepackage{amssymb}

\usepackage[figuresright]{rotating}

\begin{document}

\begin{frontmatter}



\dochead{}

\title{ Investigations of Natural Radioactivity levels and Assessment of radiological hazard of Tea samples collected from local market in Ethiopia }


\author{ Tadelech Sisay and Tilahun Tesfaye  }
 
\address{   Department of Physics, Addis Ababa University, Addis Ababa, Arat kilo, Ethiopia  }

\begin{abstract}
In this study, the activity concentration of natural ( 238 U, 232 Th, 40 K) and artificial radionuclides (137 Cs) were measurements in six different brands of tea have been collected from local market in Ethiopia (Addis Ababa city). have been analyzed by a detection system consisting of High Purity Germanium (HPGe) detector and multichannel analyzer. The average activity concentrations were nuclide M D A report 238U, 232Th and 40 K respectively. The mean values were found 3.66$ \pm$0.47, 18.28$\pm $2.4 and 582.59$ \pm $29.64 Bq/kg respectively. The activity concentration of 40K is increasing faster than the other NORM for Tea. In order to evaluate the radiological hazard of the natural radioactivity, radium equivalent (Raeq ), the external hazard index (Hex), internal hazrd index (Hin) , the total absorbed dose rate (D), the annual effective dose equivalent (AEDE) have been calculated and compared with the international acceptable value.
\end{abstract}


\end{frontmatter}


\section{ Introduction}
Radioactivity is a part of the physical environment. The main cause for the background radiation at the Earth surface are radionuclides whose origins of primordial ( 40 K, 238 U and 232Th), cosmogenic ( 3 H, 7 Be,12C,22Na) and anthropogenic ( 137 Cs,3790Sr,85Kr) types of radionuclides. It exist in the soil, rocks, plants, sand, water and air. The natural radioactivity in the environment is the main source of radiation exposure for humans. As studies suggest about 87percent of the radiation dose received by mankind in the human environment is due to natural primordial and cosmogenic radiation sources, and the remaining is due to anthropogenic radiation  [6]. Tea plants may be subjected to direct and indirect contamination of various radionuclides, such as 238U, 232Th and 40Ra. These radionuclides can be distributed in different parts of the plants according to the chemical characteristics and parameters of the plants and soil[5]. For assessing the effects of radiation exposure due to both natural and artificial radioactivity, part of which dealing with radionuclides of fallout origin. 137Cs 
and 90Sr is reported by [4]. It is crucial to estimate the ingestion dose to the public. The ingestion dose above permissible level is very harmful for human being [3]. The aim of this study is to determine natural radioactivity levels ( 238U, 232Th, 40K) and artificial (137Cs) radioactivity levels in tea samples collected from local market in Ethiopia. Also, the average radium equivalent activity (Raeq), the external hazard index (H ex ), the total absorbed dose rate (D),the annual effective dose equivalent (AEDE) which will be defined later have been calculated and compared with the results in literature.
\section{ Material and methods}
Six samples of different brands of tea have been collected from the local market in Ethiopia (Addis Ababa city) during the year of the experiment. In order to remove moisture, tea samples were dried in a drying oven at 1500$^{0}C$ until constant mass was obtained. Then samples were sealed in standardized Marinelli beaker of volume 0.5littre. Samples were kept for 30 days for secular equilibrium with 226 Ra decay products before the measurements. In this study, gamma spectroscopy analysis system was used to determine activity concentrations and radiological effects. The system consisted of a coaxial p-type high-purity germanium detector that was linked to a multichannel analyzer consisting of an analog-to-digital converter and  Genie 2000 multichannel analyzer software of 8192 channels. The spectra of all samples were perfectly analyzed by using Genie-2000 spectra analysis software to calculate the concentrations of 238U, 232Th and 40 K[2]. The spectra were analyzed for energies of daughter nuclides [ 214 Pb (295.2keV), 214 Pb (351.9 keV), 214 Bi (609.3 keV), 214 Bi (1120.2 keV), 214 Bi (1764.4 keV)] and [212 Pb (238.6 keV), 208 Tl (583.1 keV), 228 Ac (911.2 keV), 228 Ac (968.9 keV)] respectively were recorded. The activity concentrations of 40 K were determined directly by measurement of the gamma-ray transitions at 1460.8 keV.
\subsection{ Activity concentration}
The activity concentration of each radionuclides in the sample was determined by using the net count rates (cps) for the same counting time under the selected photo peaks, weight of the sample, the photo peak efficiency, and the gamma intensity at a specific  energy as given by the following equation[1].
\begin{equation}
 A=\frac{N}{P\gamma \times \varepsilon\times W }
\end{equation}
where,
N = Net counts per second (c.p.s) = (Sample
c.p.s) - (Background c.p.s)
$P_{\gamma}$= Transition probability of gamma ray
$ \varepsilon$= Efficiency in percent
W = Weight of the sample in kg. 
\subsection{Radiological effects}  
Since, the distribution of 238U , 232Th and 40K in the environment is not uniform so that with respect  to the radiation, the radiological effect of the radioactivity can be measured via absorbed dose(nGy/h), The annual effective dose Equivalent. radium equivalent activity (in Bq/Kg), and external and radiation hazard indices.
\subsection{Absorbed Dose(D(nGy/h))} 
The assessment of radiological hazard due to expo- sure of external terrestrial gamma-ray radiation from 226Ra ( 238 U ), 232Th and 40 K in plant, food, rocks, soils and building materials can also be measured using Absorbed Dose Rate in Air (ADRA) at about 1m above the ground. It is computed based on the following equation 3 [7].
\begin{equation}
 D = 0.427ARa + 0.623AT h + 0.043AK 
\end{equation}
Where the dose rate, D is in nGy/h and A stands for activity in Bq/Kg for U-238, Th-232 and K-40. This dose rate indicates the received dose from radiation emitted by radionuclides in environmental materials.
\subsection{ The Annual Effective Dose Equivalent (AEDE)} 
The annual effective dose Equivalent (AEDE) was calculated by using equation 3.
\begin{equation}
 AEDE(\mu Sv/y) =D  xDCFxOFxT
\end{equation}
where D is absorbed dose rate (nGy/ h), DCF is dose conversion factor (0.7 Sv/Gy) , OF is occupancy factor (0.2), T is the time (8760 h/y)[7].
\subsection{Radium Equivalent Calculation (Raeq)}
The specific activity of materials containing different  amounts of 238U, 232 Th and 40 K according to [1] was calculated.
\begin{equation}
 Req=ARa +1.43ATh+0.077AK
\end{equation}
 where ARa , ATh and AK are the activity
concentrations of 226 Ra, 232 Th and 40 K in Bq/kg 
respectively. The permissible maximum value of
the radium equivalent activity is 370 Bq/kg.
\subsection{ External (Hex ) and Internal (Hin ) Hazard Index}
The natural radioactivity of tea can be estimated using internal and external radiation indices based on the following  expressions[7].
\begin{equation}
 H ex = (A Ra /370) + (A Th /259) + (A K /4810)
\end{equation}
\begin{equation}
 H in = (A Ra /185) + (A Th /259) + (A K /4810)
\end{equation}
For safe use of material in the tea sample, the external index (H ex ) and internal
index (H in ) should be each less than unity.
Where: ARa, ATh and AK are the activity concentration in Bq/kg of
Ra-226, Th-232 and K-40 respectively.
\section{  Results and Discussion}
The results of activity concentrations in the Six samples of different brands of tea have been collected from the local market in Ethiopia (Addis Ababa city) are gives in table 1 for the natural radionuclides of  238 U, 232 Th and 40 K. Radium equivalent activity (Raeq ), external hazard index(Hex), internal hazards index(Hin), absorbed dose rates (D), annual effective doses equivalent(AEDE) are given in table 2\\
Table 1. Radioactivity concentration of 238 U, 232 Th and 40K in tea sample Bq/kg
\begin{center}
 \begin{tabular}{|l|p{2cm}|p{1.5cm}|l|}
 \hline
Sample\\ ID. & $^{226}$Ra  & $^{232}$Th & $^{40}$K \\
 \hline
 Tea1&3.78$ \pm 0.46 $& 19.59$ \pm 2.5 $& 565.77$\pm 23.69 $\\
 \hline
 Tea2& 2.88$ \pm 0.98$&  15.55$ \pm 2.1$& 634.11$ \pm 32.21$\\
 \hline
Tea 3& 3.66$ \pm 0.48$& 18.53$ \pm 2.59$& 518.93$ \pm 28.18$\\
 \hline
 Tea4& 3.91$ \pm 0.49$ & 18.27$ \pm 2.6$ & 573.00$ \pm 31.08 $ \\
 \hline
 Tea5& 3.67$ \pm 0.52$&18.27$ \pm 2.49$& 624.81 $ \pm 32.77$ \\
 \hline
 Tea6&  4.06$ \pm 0.46$& 19.48$ \pm 2.4$&  578.93$ \pm 29.88$\\
 \hline
  Average &  3.66$\pm 0.47 $ & 18.28$ \pm 2.4$&  582.59$ \pm 29.64$\\
\hline
  World\\ AVG &  35 & 40&  400\\
 \hline
 \end{tabular}
\end{center}
The activity concentration of the 238U in tea samples
ranged from 3.66$ \pm$0.48 Bq/kg to 4.06$ \pm$0.46 Bq/kg with
an average value of 3.66$ \pm$0.47Bq/kg. The estimated average values of 238U in
 these study are lower than the recommended maximum value of 35 Bq/kg[7].
The activity concentration of the 232 Th in tea samples
ranged from 15.55$ \pm$2.1Bq/kg to 19.59 $ \pm$2.5 Bq/kg with
an average value 18.28$ \pm$6.9 Bq/kg. The obtained value
of 232Th is lower than the acceptable value of concentration for 232 Th 40Bq/kg[7]. 
The activity concentration
of the 40 K in tea samples varied from 518.93$ \pm$28.18
Bq/kg to 634.11$ \pm$32.21 Bq/kg with an average value of
582.59$ \pm$29.62 Bq/kg. The values are greater than the
allowable value 400 Bq/kg[7].137 Cs does not exist in
tea sample naturally. The values of radium equivalent activity( Raeq) in tea samples varied from 73.95 Bq/kg to 77.87 Bq/kg with an average of 75.53 Bq/kg. The estimated average values of Raeq which is far below the internationally accepted value 370 Bq/kg[7]. The absorbed dose rates due to these radioactive nuclides in tea samples have been found to vary from 35.42 nGy/h to 39.81 nGy/h with an average value of 38.02 nGy/h. These measured values are less than the world average value 60 nGy/h. Annual effective dose equivalent have been calculated from 43.43 $ \mu$Sv/y to 48.82 $ \mu$Sv/y with an average value 46.62 $ \mu$Sv/y  respectively which is less than the annual dose limit 1mSv. The values of external hazard index, H ex range from 0.18 to 0.20 and average value was
found to be 0.19 for the tea samples. The maximum value of H ex must be less than unity. All values estimated of Hex in these study are lower than unity and the values of internal hazard index, Hin range from 0.18 to 0.21 and average value was found to be 0.19 for the tea samples. The maximum value of Hin must be less than unity. All values estimated of Hin in these studies are lower than unity. The values were less than unity in all
the samples that indicate the non-hazardous value for human being. Artificial radionuclide like 137 Cs was not found in any samples in these studies.\\
Table  2 Radium equivalent (Bq/kg),  external hazard
index, internal hazard index, absorbed dose rates(nGy/h), annual effective doses equivalent ($ \mu$Sv/y) in tea samples.
\begin{center}
 \begin{tabular}{|l|p{1.4cm}|p{1cm}|l|l||l|}
 \hline
Sample ID.  &  Raeq (Bq/kg)& Hin  & H ex  &D&ADE\\
 \hline
 Tea1&75.35& 0.2 &0.20 & 38.14& 46.77 \\
 \hline
 Tea2&73.95&   0.20&  0.19& 38.18&46.82\\
 \hline
Tea3&75.38& 0.18&  0.18& 35.42&43.43\\
 \hline
 Tea4& 74.16& 0.20 & 0.19& 37.84& 46.40\\
 \hline
 Tea5& 77.87 & 0.20 & 0.20& 39.81& 48.82\\
 \hline
 Tea6&76.49 & 0.21 & 0.20& 38.76&47.53\\
 \hline
  Average &  75.53& 0.19& 0.19& 38.02&46.62\\
\hline
  World AVG & 370& 1 & 1&60& 1\\
 \hline
 \end{tabular}
\end{center}
\section{  Conclusion}
This work to investigate the activity concentration
of natural and artificial radionuclides in Six samples of different brands of tea have been collected from the local market in Ethiopia (Addis Ababa city). The activities
concentration and the radiological hazard were measured from during the year of the experiment. The results of the study led to the following conclusions. The mean activity concentration of 238 U, 232 Th and 40 K were indented with average values of 3.66$ \pm$0.47 Bq/kg, 18.28$ \pm$2.4 Bq/kg and 582.59$ \pm$29.64 Bq/kg respectively. Annual effective dose associated with the tea samples have been found from 43.43 $ \mu$Sv/y to 48.82 $ \mu$Sv/y with an average value 46.62 $ \mu$Sv/y respectively
Which is less than   the effective dose limit 1 mSv. Values of radium equivalent activity in all samples are less than the permissible maximum value of the radium equivalent activity which is 370 Bq/kg according to UNSCEAR 2000 report. The obtained mean values of internal and external hazard indices for different tea samples are less than unity. No artificial radionuclide was found in any of the tea samples. Most of the values were less than with other study in the world. So, tea consumption in Ethiopia is non-hazardous for public health.
\section*{  Acknowledgments}
The authors are thankful to Pro.A. K. Chaubey for his valuable comments and the chairman of Physics department, Addis Ababa University for the interested in present work
\section*{ References}
[1] Beretka, J. and P. J. Mathew. 1985. Natural Radioactivity of Australian Building Materials, Industrial Wastes and By-products. Health Physics; 48: 87-95.
\newline

[2] Hailu Geremew, A. K. Chaubey and Birhanu Turi(2019) Measurements of Natural Radioactivity Levels and Associated Health Hazard Indices in Some Portland Types of cement, Ethiopia, International Journal of Scientific Research and Engineering Development-â€“ Volume 2 Issue 6,2581-7175.
\newline 
[3]  Khatun, R., Saadat, A. H. M., Ahasan, M. M., Ak- ter, S., 2013. Assessment of Natural Radioactivity and Radiation Hazard in Soil Samples of Rajbari District of Bangladesh. 2: 1-8.
\newline

[4] Lalit BY, Ramachandran TV, Rajan S (1983) Strontium- 90 and Caesium-137 in Indian tea. Radiat Environ Bio-phys 22:75-83
\newline

[5] Mst. Najnin Aktar , Suranjan Kumar Das, Selina Yeasmin , M. M. Mahfuz Siraz and A. F. M. Miza- nur Rahman(2018) MEASUREMENT OF RADIOAC- TIVITY AND ASSESSMENT OF RADIOLOGICAL HAZARD OF TEA SAMPLES COLLECTED FROM LOCAL MARKET IN BANGLADESH.J. Bangladesh Acad. Sci., Vol. 42, No. 2, 171-176.
\newline 
[6] UNSCEAR (1993). Sources, effects and Risk of Ionizing Radiation. New York: UN.
\newline

[7] UNSCEAR (2000). Sources and effects of ionization radiation, report to the general assembly, with scientific annexes. New York: UN.

\label{}




\nocite{*}
\bibliographystyle{elsarticle-num}
\bibliography{jos}







\end{document}